\documentclass[12pt]{amsart}
\usepackage{amsmath}
\usepackage{amsthm}
\usepackage{amsfonts}
\usepackage{graphicx}
\usepackage{amssymb} 
\newenvironment{nouppercase}{
  
  \renewcommand{\uppercasenonmath}[1]{}}{}

\begin{document}

\title[Gauge-compensating transformations for boosted...membranes and...]
{Gauge-compensating transformations for boosted axially symmetric membranes and light-cone reductions}
\author[Jens Hoppe]{Jens Hoppe}
\address{Braunschweig University, Germany 
\&
IHES, France}
\email{jens.r.hoppe@gmail.com}

\begin{abstract}
Some explicit examples are given for orthonormal-gauge compensating transformations and explict reductions/forms of (boosted) axially symmetric membrane solutions.
\end{abstract}

\begin{nouppercase}
\maketitle
\end{nouppercase}
\thispagestyle{empty}
\noindent
As mentioned in \cite{1}, describing axially symmetric relativistic membranes in 4-dimensional space-time by
\begin{equation}\label{eq1} 
x^{\mu}(t,\varphi,\theta) = 
\begin{pmatrix}
t\\ r(t,\varphi)\cos \theta \\ r(t,\varphi)\sin \theta \\ z(t,\varphi)
\end{pmatrix},
\end{equation}
the invariance of (generically each solution giving a minimal $\mathfrak{M}_3 \subset \mathbb{R}^{1,3}$) 
\begin{equation}\label{eq2}
\dot{r}r' +\dot{z}z'=0,\qquad \dot{r}^2 + \dot{z}^2 + r^2(r'^2 + z'^2)=1
\end{equation}
under (field-dependent) boosts in the $z$-direction,
\begin{equation}\label{eq3}
\begin{split}
t\rightarrow \tilde{t} & = t \cosh \gamma + z(t,\varphi)\sinh \gamma = ct + sz \\
z\rightarrow \tilde{z} & = z \cosh \gamma + t\sinh \gamma = cz + st 
\end{split}
\end{equation}
requires the gauge-compensating reparametrization
\begin{equation}\label{eq4}
\varphi\rightarrow \tilde{\varphi} = \varphi \cosh \gamma + v(t,\varphi)\sinh \gamma = c\varphi + sv
\end{equation}
where $v(t,\varphi)$, due to \eqref{eq2} / if $( \dot{r},\dot{z} )$
and $(r', z')$ are linear independent / implying 
\begin{equation}\label{eq5}
\ddot{z} = (r^2 z')' = 0,
\end{equation}
is consistently defined (up to a constant) by
\begin{equation}\label{eq6}
v' = \dot{z}, \qquad \dot{v} = r^2z' ;
\end{equation}
(the conserved density $\rho(\varphi) := \frac{r\sqrt{r'^2 + z'^2}}{\sqrt{1-\dot{r}^2-\dot{z}^2}}$ is, by choice of $\varphi$, taken to be $=1$, just as $\tilde{\rho}(\tilde{\varphi})$, as a consequence of \eqref{eq4}/\eqref{eq6}, analogously defined via $\tilde{z} = \tilde{z}(\tilde{t}, \tilde{\varphi})$ and $\tilde{r}(\tilde{t},\tilde{\varphi}) := r(t, \varphi)$) and itself transforms according to 
\begin{equation}\label{eq7}
v\rightarrow \tilde{v} = v \cosh \gamma + \varphi\sinh \gamma = cv + s \varphi.
\end{equation}
As it is of course instructive to consider some explicit example, let us first look 
at
\begin{equation}\label{eq8}
\begin{split}
r(t,\varphi) & = \frac{|t|}{\sqrt{2}}\bigg(\sqrt{1+8\frac{\varphi^2}{t^4}}-1\bigg)^\frac{1}{2} = tR\big(y:=\sqrt{8}\frac{\varphi}{t^2}\big)\\[0.15cm]
z(t,\varphi) & = \frac{t}{\sqrt{2}}\big(\sqrt{1+y^2}+1\big)^\frac{1}{2} = tQ(y),  
\end{split}
\end{equation}
satisfying 
\begin{equation}\label{eq9}
t^2 + r^2 -z^2 = 0
\end{equation}
\begin{equation}\label{eq10}
R^2Q^2 = \frac{y^2}{4}, \;  Q^2 = R^2 +1, \; Q^4 = Q^2 + \frac{y^2}{4}
\end{equation}
and, via ($t>0$)
\begin{equation}\label{eq11}
\dot{r} = R - 2yR', \; r' = \frac{t}{\varphi}yR', \; \dot{z} = Q - 2yQ', \; z' = \frac{t}{\varphi}yQ',
\end{equation}
\begin{equation}\label{eq12}
y(R^2)' = \frac{2R^2(R^2+1)}{2R^2+1}, \;\; y(Q^2)' = \frac{2Q^2(Q^2-1)}{2Q^2-1}
\end{equation}
(from \eqref{eq2}$_1$), implying
\begin{equation}\label{eq13}
y^2 = c^2R^2(R^2+1) = c^2Q^2(Q^2-1),
\end{equation}
as well as (from \eqref{eq2}$_2$)
\begin{equation}\label{eq14}
R^2 - y(R^2)' + \frac{(R^4)'}{y} = 0 ,
\end{equation}
which fixes the integration constant $c^2$ in \eqref{eq13} to be $= 4$ (this is one possible way to actually derive the explicit form in \eqref{eq8} - which however originally \cite{2} was found in quite a different way).
Applying \eqref{eq3}/\eqref{eq4} to the above example, with \eqref{eq6}, and $v(t) = \varphi\cdot G(y)$ implying
\begin{equation}\label{eq15}
\begin{split}
G + yG' & = Q - 2yQ'\\
-2G'& = \frac{t^4}{\varphi^2}R^2Q' = 8\frac{R^2Q'}{y^2},
\end{split}
\end{equation}
i.e. (in the last step using \eqref{eq14})
\begin{equation}\label{eq16}
\begin{split}
G & = Q-2yQ' + \frac{4R^2Q'}{y} \\
  & = \frac{1}{\sqrt{R^2 +1}}\big( 1+R^2 - y(R^2)' + 4R^3R'\big) = \frac{1}{Q}
\end{split}
\end{equation}
one gets
\begin{equation}\label{eq17}
\begin{split}
\tilde{t} & = t(c+sQ(y)) \\
\tilde{\varphi} &= \varphi(c + \frac{s}{Q}) = \varphi \frac{(cQ + s)}{Q}
\end{split}
\end{equation}
i.e 
\begin{equation}\label{eq18}
\begin{split}
\tilde{y} & = \sqrt{8}\frac{\tilde{\varphi}}{\tilde{t}^2} = y \frac{(cQ+s)}{Q(c+sQ)^2} \\[0.15cm]
\tilde{R}(\tilde{y}) & = \frac{R(y)}{c+sQ(y)} \\[0.15cm]
\tilde{Q}(\tilde{y}) & = \frac{cQ+s}{c+sQ}. 
\end{split}
\end{equation}
While it at first seems difficult to deduce $y = K(\tilde{y})$ (in order to obtain the dependence of $\tilde{R}$ and $\tilde{Q}$ on $\tilde{y}$) one can check that 
\begin{equation}\label{eq19}
\begin{split}
(2\tilde{Q}^2-1)^2 & = \big( 2 (\frac{cQ+s}{c+sQ})^2 -1 \big)^2 \\[0.15cm]
& = 1+y^2\frac{(cQ+s)^2}{Q^2(c+sQ)^4} = 1+\tilde{y}^2 ,
\end{split}
\end{equation}
as well as (easier)
\begin{equation}\label{eq20}
y = \tilde{y} \frac{c\tilde{Q}-s}{\tilde{Q}(c-s\tilde{Q})^2};
\end{equation}
in any case, as $\mathfrak{M}_3$ given by \eqref{eq9} is obviously boost-invariant, and the orthonormal parametrization, cp. \eqref{eq8}-\eqref{eq14}, determined by \eqref{eq2} and \eqref{eq9}, $\tilde{R}$ and $\tilde{Q}$ (as functions of $\tilde{y}$) {\it have} to be the same functions than $R$ and $Q$ (as functions of $y$); but it is good to see this being implied explicitly by \eqref{eq4}.\\
While in general it will of course be extremely different to explicitly describe the finite (compared to `only infinitesimal') action of the Lorentz-boosts (i.e. inverting $\tilde{t} = ct + sz(t,\varphi)$, $\tilde{\varphi} = c\varphi + sv(t,\varphi)$ to obtain $\tilde{z} = cz(t,\varphi)+st$ as a function of $\tilde{t}$ and $\tilde{\varphi}$ - which then, together with $\tilde{r}(\tilde{t},\tilde{\varphi}) := r(t,\varphi)$ will satisfy \eqref{eq2}, with $\dot{•}$ and $'$ being derivatives with respect to $\tilde{t}$ and $\tilde{\varphi}$, and the boosted $\mathfrak{M}_3$ having $\tilde{x}^{\circ}(\tilde{t},\tilde{\varphi},\tilde{\theta}) = \tilde{t}$) there is one (in several aspects quite interesting) simple case, where it can easily be done, namely when applying \eqref{eq3}-\eqref{eq6} to a {\it static} axially symmetric minimal surface $\Sigma_2$, resp. $\Sigma_2 \times \mathbb{R}$, i.e. the Catenoid - which in the parametrization $\rho(\varphi)=1$ reads
\begin{equation}\label{eq21}
x^{\mu}(t, \varphi,\theta) = 
\begin{pmatrix}
t \\ \cosh z(\varphi) \cos \theta \\ \cosh z(\varphi) \sin \theta\\ z(\varphi)
\end{pmatrix},
\end{equation}
$r^2(r'^2 + z'^2) = \cosh^2z(\varphi)z'^2(1+\sinh^2z(\varphi)) = 1$ implying
\begin{equation}\label{eq22}
z'(\varphi) = \frac{1}{\cosh^2 z(\varphi)},
\end{equation}
i.e.
\begin{equation}\label{eq23}
\varphi(z) = \frac{1}{2}(\sinh z \cosh z + z)
\end{equation}
($\frac{d\varphi}{dz} = \cosh^2z > 0$, i.e.\eqref{eq23} uniquely defining the inverse function $z(\varphi)$, up to a constant shift).\\
This (due to $\dot{r} = 0 =\dot{z}$) being a case where \eqref{eq2} does a priori not necessarily imply the second order `membrane-equations'
\begin{equation}\label{eq24}
\ddot{z} = (r^2z')', \qquad \ddot{r} = (r^2r')' - r(r'^2+z'^2) ,
\end{equation}
it is better to check:
$r^2z' = 1$ implies the first part of (24), while $(r^2r')' - rr'^2 = (\sinh z)' - \frac{\sinh^2 z}{\cosh^3 z} = \frac{1}{\\cosh^3 z} = rz'^2$ verifies the second.
\begin{equation}\label{eq25}
v' = \dot{z} = 0, \qquad \dot{v} = r^2z' = 1
\end{equation}
(here, in particular, it pays to have chosen the unusual parametrization \eqref{eq21}) implies $v(t,\varphi) = t-t_0$, i.e. (choosing $t_0 = 0$)
\begin{equation}\label{eq26}
\tilde{t} = ct + sz(\varphi), \qquad \tilde{\varphi} = c\varphi + st
\end{equation}
being simple enough to conclude/see that
\begin{equation}\label{eq27}
\begin{split}
s\tilde{t} - c\tilde{\varphi} & = s^2z(\varphi) - c^2\varphi\\
& = s^2z - \frac{c^2}{2}(z + \sinh z \cosh z)
\end{split}
\end{equation}
(due to the right-hand side having strictly negative $\varphi$-derivative, $=\frac{s^2}{\cosh^2 \varphi} - c^2 <0$) gives $\varphi$ as a function of $s\tilde{t}-c\tilde{\varphi}$, resp.
\begin{equation}\label{eq28}
z(\varphi) = k(s\tilde{t}-c\tilde{\varphi}),
\end{equation}
hence
\begin{equation}\label{eq29}
\begin{split}
\tilde{z} & = cz(\varphi) + st\\
& = ck(s\tilde{t} - c\tilde{\varphi}) + \frac{s}{c}(\tilde{t}-sk)\\
& = \frac{1}{c}(k(s\tilde{t} - c\tilde{\varphi})+s\tilde{t}),
\end{split}
\end{equation}
\begin{equation}\label{eq30}
\tilde{r}  = \cosh z(\varphi)
= \cosh(k(s\tilde{t} - c\tilde{\varphi}))
= \cosh (c\tilde{z} - s\tilde{t}),
\end{equation}
the last expression (giving the obvious boosted level set version of $r = \cosh z$) also corresponding to the reduction of\footnote{many thanks to V.Sokolov for several related discussions}
\begin{equation}\label{eq31}
\ddot{R} = R(RR')'
\end{equation}
via the Ansatz $R(\tau,\mu) = y(x = \mu + \nu\tau)$, giving $(\zeta = t-z = \zeta(\tau,\mu))$
\begin{equation}\label{eq32}
\begin{array}{l}
y''(y^2-\nu^2) + yy'^2  = 0, \quad y' = \frac{e}{\sqrt{y^2 - \nu^2}},\\[0.25cm]
2e(x-x_0) = y\sqrt{y^2-\nu^2} - \nu^2 \text{arccosh} \frac{y}{\nu}\\[0.25cm]
R(\mu + \nu\tau) = \nu \cosh(\frac{\zeta}{e\nu} - \frac{e\tau}{2\nu} = \alpha\tau + \beta\zeta)\\[0.25cm]
2\alpha\beta\nu^2 = -1,
\end{array} 
\end{equation}
resp.
\begin{equation}\label{eq33}
\begin{split}
r(t,z)  & = \nu \cosh (\gamma t + \delta z)\\[0.15cm]
\nu^2(\delta^2 - \gamma^2) & = +1 .
\end{split}
\end{equation}
Another reduction,
\begin{equation}\label{eq34}
R(\tau, \mu) = \tau^a f(\tau^b \mu)_{a+b+1 = 0},
\end{equation}
considered already in \cite{3} (and linear solutions, for small integer values of $a$, mentioned in a grant appication in 2021) reduces \eqref{eq31}, via
\begin{equation}\label{eq35}
f(s) = sg(\ln s =: u), \; f' = g +g', \; sf'' = g'+g'',
\end{equation}
to the ODE
\begin{equation}\label{eq36}
2g + 3(a+1)g' + (a+1)^2 g'' = g (g''g + g'^2 + g^2 +3gg'),
\end{equation}
for $a=-1$ reducing to a linear equation for $g^2$, whose general solution is
\begin{equation}\label{eq37}
g^2(u) = 2\{ 1 + Ae^{-u} + B e^{-2u}\},
\end{equation}
and for general $a$, via $G(g) := g'$, giving the first order ODE
\begin{equation}\label{eq38}
2g + 3(a+1)G + (a+1)^2 GG' = g(gGG' + G^2 + g^2 +3gG)
\end{equation}
which has linear solutions
\begin{equation}\label{eq39}
G(g) = \gamma g + \delta,
\end{equation}
for\footnote{noticed independently by V.Sokolov} $a = -1,0,1,2,3$,
giving 
\begin{equation}\label{eq40} 
\begin{split}
g(u) & = D^2e^{\gamma u} - \frac{\delta}{\gamma}\\[0.15cm]
f(s) & = s(D^2 s^{\gamma} - \frac{\delta}{\gamma}) = D^2 s^{\gamma + 1} - \frac{\delta}{\gamma}\cdot s,
\end{split}
\end{equation}
where $\gamma$ and $\delta$ take the following values for the above mentioned integer values of $a$ :
\begin{equation}\label{eq41} 
\begin{split}
a = -1 & : \; \gamma = -1,\, \delta = \pm \sqrt{2}\\
a = 0 & : \; \gamma = -1,\, \delta = 0\\
a = 1 & : \; \gamma = -1,\, \delta = 0 \, \text{or} \, \gamma = -\frac{1}{2},\, \delta = 0 \\
a = 2 & : \; \gamma = -1,\, \delta = \pm \sqrt{2}\\
a = 3 & : \; \gamma = -\frac{1}{2},\, \delta = 0,
\end{split}
\end{equation}
corresponding to solutions (slightly generalizing the involved constants) 
\begin{equation}\label{eq42} 
\begin{split}
R^{b=0} & = \frac{1}{\tau} (\varepsilon \pm \sqrt{2}\mu) = \hat{R}^{(0)}, \, \tilde{R}^{b=-1} = \varepsilon, \,\tilde{R}^{b=-2} = \varepsilon \tau \\ 
R^{b=-2} & = \tau (\varepsilon  \sqrt{\mu/\tau^2}) \rightarrow \hat{R} = \sqrt{\mu}(\varepsilon + q\tau) \\ 
R^{b=-3} & = \tau^2 (\varepsilon \pm \sqrt{2}\mu/\tau^3)  \\
R^{b=-4} & = \tau^3 (\varepsilon  \sqrt{\mu/\tau^4}) \rightarrow \hat{R} = \sqrt{\mu}(\varepsilon + q \tau). 
\end{split}
\end{equation}
Noting that the $\tilde{R}$ - solution are `fake'/`singular' ({\it not} corresponding to minimal $\mathfrak{M}_3$'s), and the $\hat{R}$'s being trivial generalizations, e.g. in level-set form:
\begin{equation}\label{eq43} 
R^2 = (t-z)(2\varepsilon + q(t+z)) - \frac{(\varepsilon + \frac{1}{2}q(t+z))^6}{2},
\end{equation}
of, going back to \cite{3} (see also \cite{4},\cite{5} and eqs. 21/31 of \cite{6}), previously well - discussed solutions, let us look at several interesting aspects of
\begin{equation}\label{eq44} 
R = \pm \sqrt{2} \frac{\mu}{\tau} + \varepsilon \tau^2
\end{equation}
whose level set description (of the corresponding $\mathfrak{M}_3$) follows from calculating $\zeta = t -z$ from
\begin{equation}\label{eq45} 
\begin{split}
\zeta' & = \dot{R}R' = (\mp \sqrt{2}\frac{\mu}{\tau^2} + 2\varepsilon\tau)(\pm \frac{\sqrt{2}}{\tau}) \\[0.15cm]
\dot{\zeta} & = \frac{1}{2}(\dot{R}^2 + R^2R'^2) = \frac{1}{2}\lbrace (\mp \sqrt{2} \frac{\mu}{\tau^2} + 2 \varepsilon \tau)^2 + (2\frac{\mu}{\tau^2} \pm \sqrt{2}\varepsilon \tau)^2 \rbrace,
\end{split}
\end{equation}
i.e. (up to a constant $\zeta_0$)
\begin{equation}\label{eq46} 
\zeta = \varepsilon^2 \tau^3 - \frac{\mu^2}{\tau^3} \pm 2\sqrt{2}\mu \varepsilon,
\end{equation}
hence
\begin{equation}\label{eq47} 
R^2 + 2\zeta\tau - 6 \varepsilon R \tau^2 + 3 \varepsilon^2 \tau^4 = 0
\end{equation}
resp. (with $\varepsilon = 4C$)
\begin{equation}\label{eq48} 
u:=(t^2 + x^2 +y^2 - z^2) - 6C \sqrt{x^2 + y^2}(t+z)^2 + 3C^2(t+z)^4 = 0, 
\end{equation}
for $C= 0$ giving the hyperboloid discussed above (cp. eq's \eqref{eq8}-\eqref{eq20}), a time-like minimal $\mathfrak{M}_3$, (i.e. valid membrane solution), except that the points $z = \pm t$ generate light-like lines. 
For $C \neq 0$, it remains time-like (always except the light-like lines at $R=0$), as can be seen by calculating the square of the Minkowski-gradient of the left-hand side of \eqref{eq48}, $(\cdot \frac{1}{4})$, to be  
\begin{equation}\label{eq49} 
\begin{array}{l}
\big(t+6C(t+z)((t+z)^2C-R)\big)^2 - x^2 - y^2 \\[0.15cm]
\qquad \qquad - 9C^2(t+z)^4 + 6R(t+z)^2C \\[0.15cm]
\qquad \qquad - \big( z -C(t+z)((t+z)^2C - R) \big)^2 \\[0.15cm]
 = t^2 - R^2 - z^2 - 9C^2(t+z)^4 + 6R(t+z)^2C \\[0.15cm]
\; \; \; + 12C(t+z)^2((t+z)^2C-R)  
\end{array}
\end{equation}
which on $\mathfrak{M}_3$ (i.e using \eqref{eq48} to replace $t^2 - R^2 - z^2 = (t^2 + R^2 - z^2) - 2R^2$) becomes \footnote{That the square of the gradient of u equals $-8R^2+4u$    also significantly simplifies the proof that the set of regular points where u vanishes defines a zero-mean-curvature 3-manifold; furthermore, letting the last term of (48) and the coefficients of $R^2$ and R be arbitrary functions of (t+z) - say 2g,h, and 2f ( as well as letting d:=D-2 be arbitrary, rather than 2) one finds that $8(M-1)bRf-8(2-Mh)bR^2$ and 1/2 of the scalar product of the Minkowski-gradient of u with the derivative of $4[u-bR^2+2R(f'(t+z)-f-hf)-(2g+f^2-2g'(t+z))]$ (can only) matchfor h=1, M=2 (implying b=2) as well as f and g being quadratic and quartic, as in (48)}
\begin{equation}\label{eq50} 
-2R^2 + 12C(R-(t+z)^2C)[(t+z)^2-(t+z)^2] = -2R^2 \leq 0.
\end{equation}
To view \eqref{eq48} as a t(ime)-evolution of 2 dimensional axially symmetric surfaces i.e. time-dependent planar curves given, for each $t$, by some relation between $r := \sqrt{x^2+y^2}$ and $z$, one may either note that (from \eqref{eq48})
\begin{equation}\label{eq51} 
\frac{r}{(t+z)^2}\big( 1 \pm \sqrt{\frac{2}{3} + \frac{z^2-t^2}{3r^3}}\big) = C,
\end{equation}
or use that \eqref{eq48} implies
\begin{equation}\label{eq52} 
\begin{split}
r(t,z) & = (t+z)^2 \big(3C \pm \sqrt{6C^2 + f(z;t)}  \big)\\[0.15cm]
f(z) & := \frac{z-t}{(z+t)^3};
\end{split}
\end{equation}
$f'(z) = \frac{4t-2z}{(z+t)^4}$ vanishing only at $z = 2t =: z_+$, and $f$ diverging for $z \rightarrow -t$, one can easily see that for $C < 0$ (which is simpler, as only the upper sign in \eqref{eq52} is relevant), and $t > \frac{1}{9|C|}$ (which implies that around $z_+$, $f < 3C^2$), \eqref{eq52} describes a simple closed curve - as in that case $f(z) \geq 3C^2$ is only satisfied for 
\begin{equation}\label{eq53} 
\begin{array}{l}
z \in [-\delta t - t, -t)\\[0.15cm]
3C^2 (\delta t)^3 - \delta t - 2t = 0
\end{array}
\end{equation}
(hence $\delta t \sim \frac{t^{\frac{1}{3}}}{(C^2)^{\frac{1}{3}}} $ for $ t \rightarrow \infty$); $r_{\text{max}} \lesssim \frac{t^{\frac{2}{3}}}{|C|^{\frac{1}{3}}}$
\begin{equation}\label{eq54} 
\begin{array}{l}
\frac{\partial r}{\partial z} = 2(t+z)(3C + \sqrt{6C^2 + f}) + \frac{2t - z}{\sqrt{6C^2 + f}(t+z)^2}\\[0.35cm]
\end{array}
\end{equation}
implies $\frac{\partial r}{\partial z} \searrow -\infty$ for $z \nearrow -t$ while $\frac{\partial r}{\partial z}(-t-\delta t)$ is finite, $> 0$.\\
Note that (48), for negative C (say, =-1) describes (for $t<-1/9$ one with $z>0$, for $t>1/9$ one with $z<0$, for $|t|<1/9$ two) fast moving drops with with a sharp tip pointing to the nearest infinity (the $z>0$ drop constantly shrinking, the drop with $z<0$ constantly growing); that the drops reverse their momentum at t=0 (the only time where they, at the ``infinite energy" origin, have contact with each other) may either be interpreted as ``bouncing (off one another)" or transferring momentum and energy at the origin.\\
Finally, let us go back to \eqref{eq44} and consider general `perturbations' of $R_0 = \sqrt{2}\frac{\mu}{\tau}$,
\begin{equation}\label{eq55} 
R(\tau, \mu) = R_0 + \varepsilon R_1 + \varepsilon^2 R_2 + \ldots = \sqrt{2}\frac{\mu}{\tau} + \sum^{\infty}_{n=1}R_n\varepsilon^n,
\end{equation}
in each order having to solve some (for $n \geq 2$, inhomogeneous) linear ODE,
\begin{equation}\label{eq56} 
LR_n := \tau^2 \ddot{R}_n - 2(\mu^2 R''_n + 2\mu R'_n + R_n) = F_n.
\end{equation}
$L$ being particularly simple, one has
\begin{equation}\label{eq57} 
\begin{split}
R_n & = \int \tau^{\alpha}\mu^{\beta(\alpha)}h_n(\alpha)\,d \alpha + \hat{R}_n \\[0.15cm]
\beta(\alpha) & = -\frac{1}{2} \pm \sqrt{\frac{1}{4}+\frac{\alpha^2-\alpha}{2}-1} 
\end{split}
\end{equation}
i.e. in each order the freedom to choose $h_n^{(\pm)}(\alpha)$ (for the solution of $LR^h_n = 0$; $\hat{R}_n$ a particular solution of the inhomogeneous equation \eqref{eq56}). As $F_1 \equiv 0$, 
\begin{equation}\label{eq58} 
\begin{split}
F_2 & = \tau^2 \left[ R_0(R''_1 R_1 + R'^2) + R_1(R''_0 R_1 + 2R'_0R'_1 + R_0 R''_1) \right] \\[0.15cm]
& = \sqrt{2}\tau \left[ 2\mu R ''_1 R_1 + \mu R'^2_1 + 2R_1R'_1\right] \\[0.15cm]
F_n & = \sum\tau^2 \left[ R_k(R''_l R_m + R'_lR'_m) \right]
\end{split}
\end{equation}
(where the right-hand side of $F_n$ is summed over all k+l+m=n, with $k,l,m < n $)
one way of viewing \eqref{eq44} is that $F_2 \equiv 0$ (hence the possibility to choose $R_2 = 0$) if $R'_1 = 0$; $\beta(\alpha) = 0$ implies $\alpha^2-\alpha =2$, i.e. $\alpha = 2$ (giving \eqref{eq44}), or $= -1$ (corresponding to a trivial shift of $\mu$).\\
\textbf{Acknowledgement}:
I would like to thank J.Arnlind, V.Bach, I.Bobrova, T.Damour, J.Eggers,
I.Gahramanov, R.Hernandez, G.Huisken, K.Kim,  M.Kontsevich, S.Lynch,  I.Nafiz, L.Mason, 
V.Roubtsov, V.Sokolov, and T.Turgut for discussions.


\begin{thebibliography}{11111}
\bibitem[1]{1} J.Hoppe, {\it Integrability in the dynamics of axially symmetric membranes}, arXiv:2202.06955, 2022
\bibitem[2]{2} J.Hoppe, {\it U(1) invariant minimal hypersurfaces in $\mathbb{R}^{1,3}$} Phys.Lett.B 736, 2014, 465
\bibitem[3]{3} J.Hoppe, {\it Some classical solutions of relativistic membrane equations in 4 space-time dimensions}, Phys.Lett.B 329, 1994 
\bibitem[4]{4} J.Hoppe, {\it Exact algebraic M(em)brane solutions}, 
Asian J.Math Vol.26, 2, p.253 2022
\bibitem[5]{5} J.Hoppe, {\it On the quantization of some polynomial minimal surfaces},
Phys.Lett.B 822, 2021
\bibitem[6]{6} J.Choe,J.Hoppe.T.Turgut, {\it Generating axially symmetric minimal hypersurfaces ...}, arXiv:2211.03887, 2022
\end{thebibliography}
\end{document}